\documentclass[final,5p,times,twocolumn]{elsarticle}

\journal{Nuclear Instruments and Methods in Physics Research - section A (NIM-A)}

\usepackage{subfigure}

\begin{document}

\begin{frontmatter}

\title{Skipper-CCDs: current applications and future}

\author[Fermilab,UNAM]{Cervantes-Vergara, B.A.\corref{mycorrespondingauthor}}
\cortext[mycorrespondingauthor]{Corresponding author. Email: \textit{bren.cv@ciencias.unam.mx}}
\author[Fermilab,UBA]{Perez, S.}
\author[UNAM]{D'Olivo, J.C.}
\author[Fermilab]{Estrada, J.}
\author[Microchip]{Grimm, D.J.}
\author[LBNL]{Holland, S.}
\author[Fermilab,UNC]{Sofo-Haro, M.}
\author[Microchip]{Wong, W.}

\address[Fermilab]{Fermi National Accelerator Laboratory, Batavia, IL, USA}
\address[UNAM]{Universidad Nacional Autónoma de México, Ciudad de México, México}
\address[UBA]{Universidad de Buenos Aires, Buenos Aires, Argentina}
\address[UNC]{Universidad Nacional de Córdoba, Córdoba, Argentina}
\address[LBNL]{Lawrence Berkeley National Laboratory, Berkeley, CA, USA}
\address[Microchip]{Microchip Technology Inc., Chandler, AZ, USA}

\begin{abstract}
This work briefly discusses the potential applications of the Skipper-CCD technology in astronomy and reviews its current use in dark matter and neutrino experiments. An overview of the ongoing efforts to build multi-kilogram experiments with these sensors is given, in the context of the Oscura experiment. First results from the characterization of Oscura sensors from the first 200~mm wafer-fabrication run with a new vendor are presented. The overall yield of the electron counting capability of these sensors is 71\%. A noise of 0.087~e$^-$ RMS, with 1225 samples/pix, and a dark current of $(0.031\pm 0.013)$~e$^-$/pix/day at 140~K were measured.
\end{abstract}


\end{frontmatter}

\section{Introduction}

Scientific Charge-Coupled Devices (CCDs) have been widely used in astronomy and particle physics due to their great spatial resolution and low-energy threshold. Since the successful demonstration of the electron counting capability of the Skipper-CCD, with a design made by Stephen Holland from the Lawrence Berkeley National Laboratory (LBNL)~\cite{Tiffenberg2017}, this new-generation sensor has increasingly received attention for its potential to detect low-energy signals ($\sim$eV). In astronomy, observations of faint sources as well as high-cadence searches for short duration transients could benefit from the sub-electron noise achieved with these sensors in both imaging and spectroscopy~\cite{Drlica2020, SnowmassAstro2022}. In particle physics, this technology is currently used in low-energy neutrino studies and dark matter (DM) searches. First results in these areas are very promising.

\section{State-of-the-art in Skipper-CCD applications}

\paragraph{Direct DM search} The SENSEI Collaboration~\cite{SENSEI2020} has successfully demonstrated the Skipper-CCD potential in this area by imposing world-leading constraints on DM-electron interactions in the low-mass regime using a $\sim$2~g detector. Several active and planned experiments form part of the ongoing effort to search for DM with Skipper-CCDs. The DAMIC experiment at SNOLAB has recently installed $\sim$18~g of Skipper-CCDs to probe its previously reported excess of events at low energies~\cite{DAMIC2020}. SENSEI-100, planning to install a 100~g detector at SNOLAB by the end of 2022, has undergone a partial commissioning and is currently taking data. The DAMIC-M Collaboration~\cite{DAMIC-M}, whose ultimate goal is to install a 1~kg detector of Skipper-CCDs at Laboratoire Souterrain de Modane in 2023, is now taking data with its $\sim$18~g proof-of-concept prototype. Finally, the Oscura experiment~\cite{Oscura}, currently finishing its R\&D stage, aims to deploy 10~kg of Skipper-CCDs at SNOLAB by 2026 to lead the search for low-mass dark matter particles.

\paragraph{Low-energy neutrinos} Skipper-CCDs are also promising sensors to probe new physics with reactor neutrino experiments~\cite{Fernandez-Moroni2020,Fernandez-Moroni2021}. In this regard, the CONNIE experiment~\cite{CONNIE2021} has been upgraded with Skipper-CCDs to study their performance and background at sea level, providing an experimental basis for estimating the physics potential of a Skipper-CCD reactor neutrino experiment.

\section{Towards multi-kg Skipper-CCD experiments (Oscura)}

All of the previously discussed experimental efforts aim to improve their sensitivities by lowering their low-energy backgrounds and by increasing their masses. Nowadays, g-size detectors in which tens of sensors and readout channels are involved can be built with the current technology. Increasing the mass entails the development of new ideas on sensor packaging, cryogenics and electronics. During the R\&D for Oscura, all of these areas have been addressed~\cite{Oscura}. Particularly, the development of the Oscura's readout electronics is one of the three main technical challenges that have been identified and the results from ongoing efforts will be published elsewhere.

Regarding background, Oscura's design goal is to have less than one 2e$^-$ background events for the full exposure of 30~kg-year~\cite{Oscura}. This requirement imposes constraints in the radioactive background rate (0.01 dru), in the sensors dark current rate ($\sim10^{-6}$~e$^-$/pix/day) and in the full array readout time ($<2$~hrs). Radioactive background control and the fabrication of science-grade sensors are the other two major technical challenges for Oscura. In this work we will discuss in detail the latter.

Until now, Teledyne DALSA has been the only foundry fabricating Skipper-CCDs for dark matter experiments, but they are stopping this production line. Oscura has found a new vendor, Microchip Technology Inc., and has adapted the LBNL Skipper-CCD design by Stephen Holland to the Microchip 200~mm-wafer processes. The first fabrication run at Microchip was completed at the end of summer 2021.

\subsection{Characterization of sensors from Microchip}

The Skipper-CCDs have been fabricated in 200~mm diameter high-resistivity Si wafers. In the first iteration 25 wafers were processed, and a buried-channel ion implantation dose split was done to match the DALSA process. Each sensor is 675~$\mu$m thick and has an active area of 1.9~cm $\times$ 1.6~cm. Charge can be read through four amplifiers, one in each corner.
\begin{figure}[h!]
    \centering
    \begin{subfigure}{}
    \includegraphics[height=2.8cm, width=2.8cm]{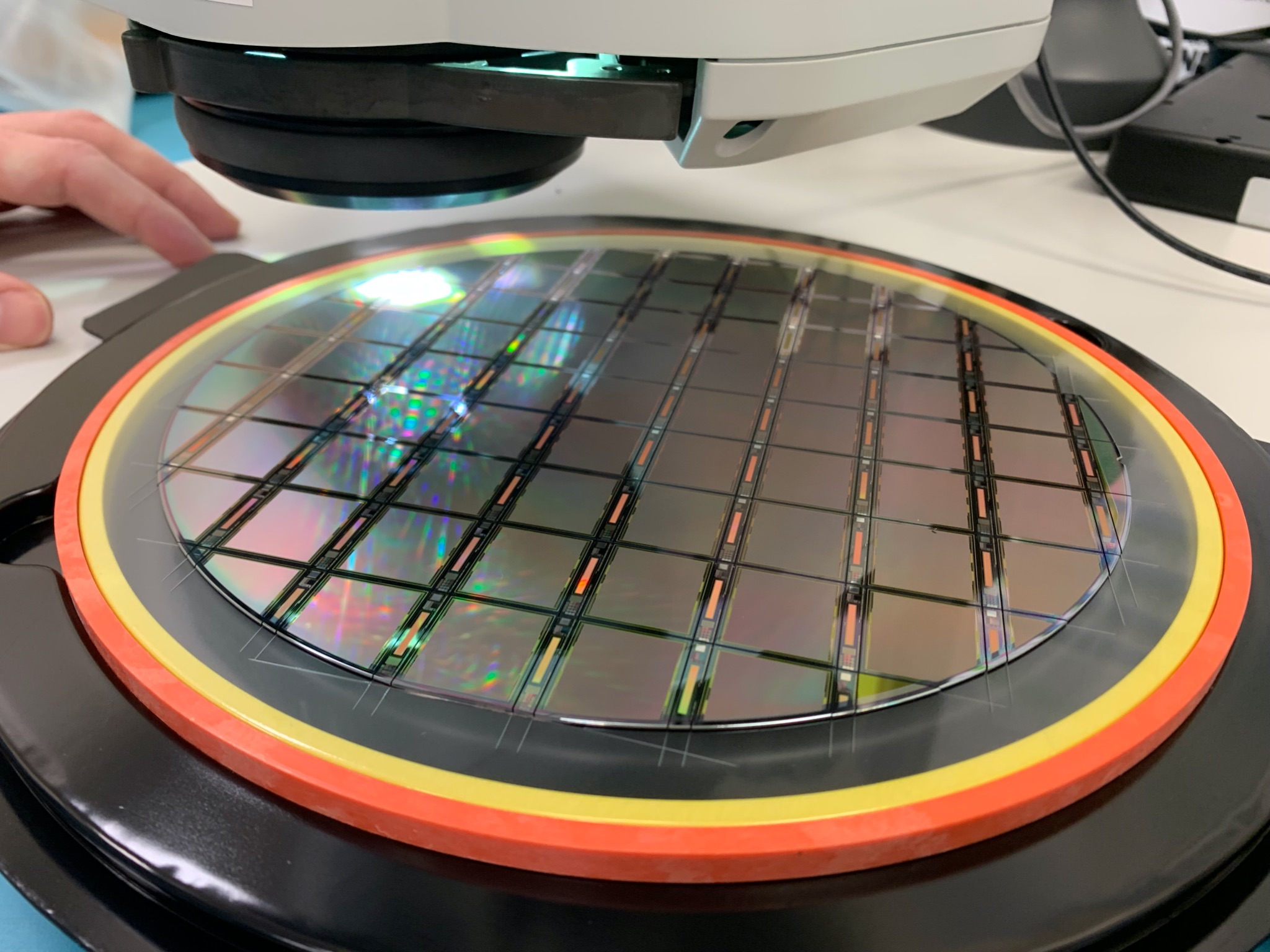}
    \end{subfigure}
    \begin{subfigure}{}
    \includegraphics[height=2.5cm]{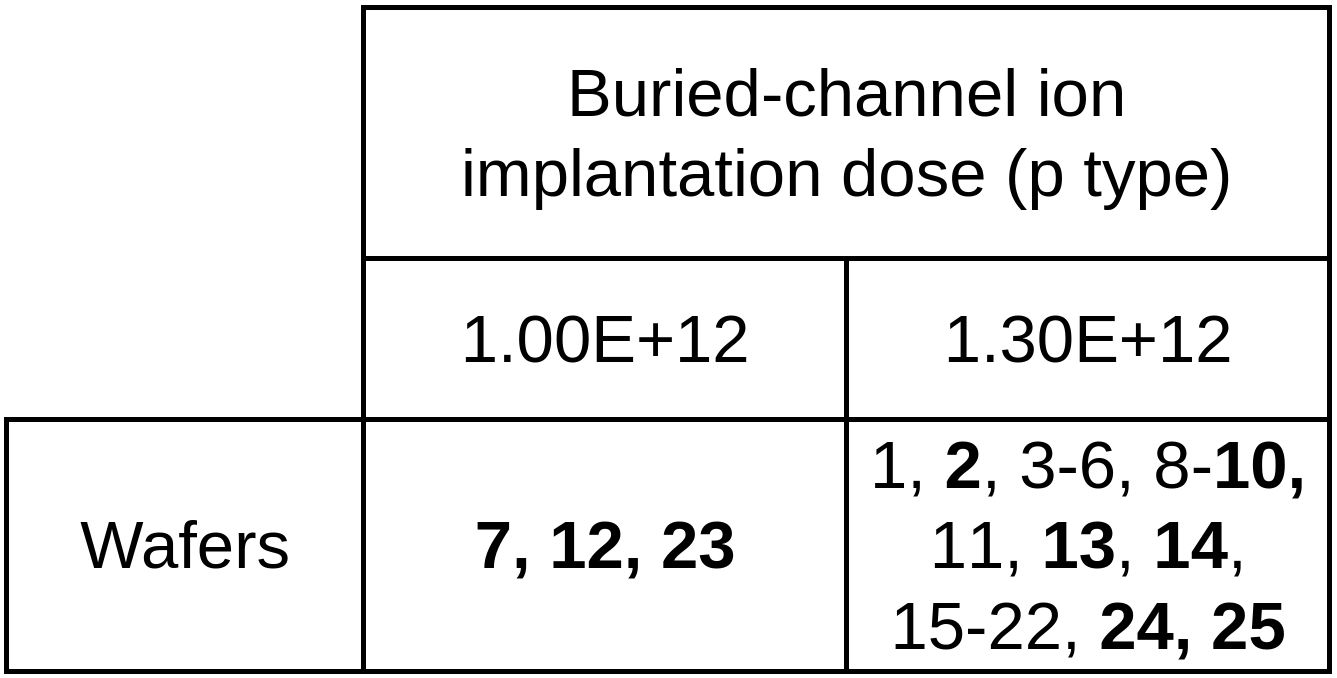}
    \end{subfigure}
    \caption{Left: 200~mm diameter wafer. Right: Corresponding buried-channel ion implantation doses for each wafer. Sensors from wafers in bold were tested.}
    \label{fig:wafers}
\end{figure}

Skipper-CCDs from 9 different wafers were packaged in Cu modules and tested in some dedicated setups installed in the Silicon Detector Facility, at the Fermi National Accelerator Laboratory (FNAL). Using a cryocooler, the sensors are operated at around 150~K. A low-threshold acquisition board (LTA)~\cite{LTA} is used for reading out the collected charge.
\begin{figure}[h!]
    \centering
    \begin{subfigure}{}
    \includegraphics[height=3cm]{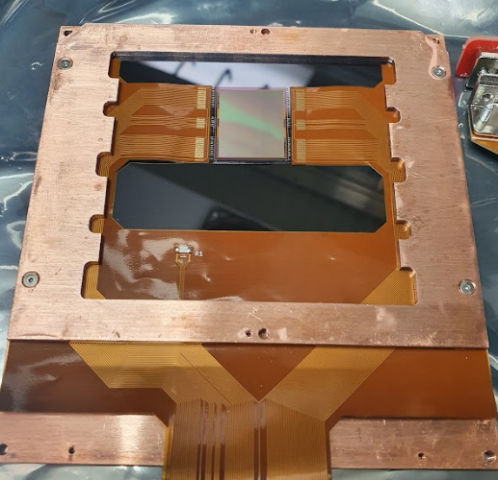}
    \end{subfigure}
    \begin{subfigure}{}
    \includegraphics[height=3cm]{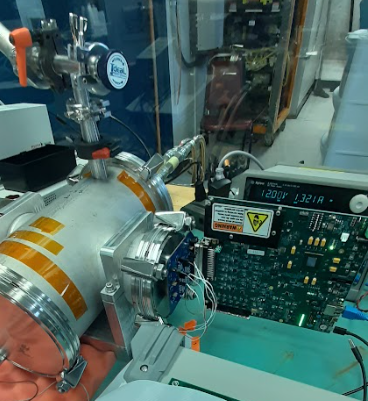}
    \end{subfigure}
    \begin{subfigure}{}
    \includegraphics[height=3cm]{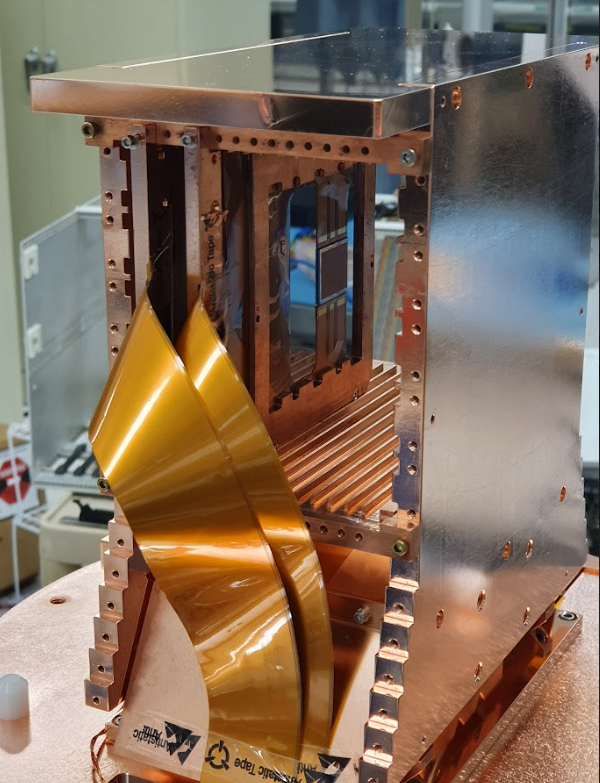}
    \end{subfigure}
    \caption{Left: Microchip Skipper-CCD in Cu module. Center and right: Setups at FNAL for testing packaged sensors.}
    \label{fig:setups}
\end{figure}

\begin{figure}[h!]
    \centering
    \begin{subfigure}{}
    \includegraphics[height=3cm]{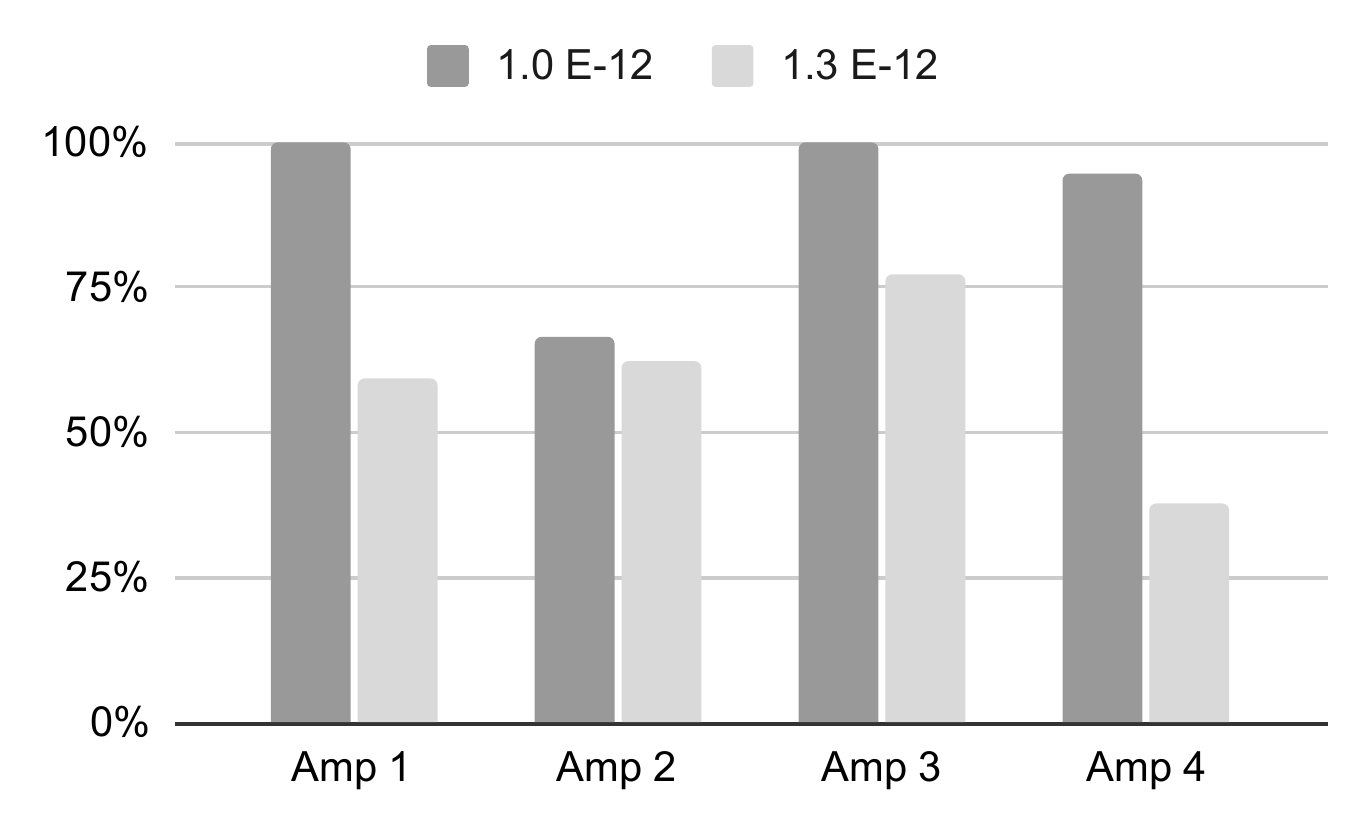}
    \end{subfigure}
    \begin{subfigure}{}
    \includegraphics[height=3.4cm]{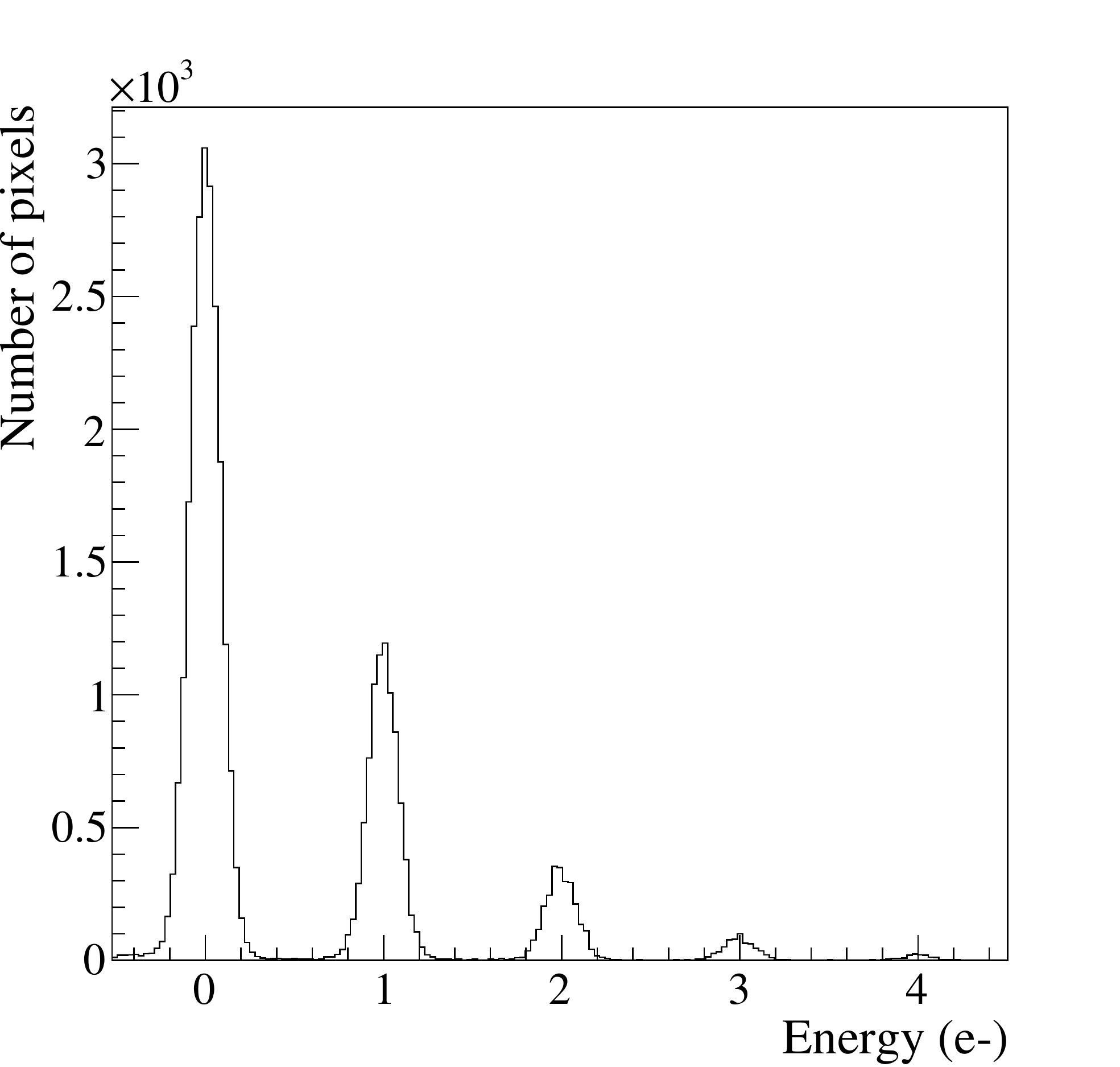}
    \end{subfigure}
    \caption{Left: Percentage of tested sensors with single-e$^-$ resolution per amplifier and buried-channel ion implantation dose. Right: Pixel charge distribution from an image with 1225 samples per pixel (with 0.087~e$^-$ RMS of noise), taken with one of the Microchip sensors.}
    \label{fig:epeaks}
\end{figure}

After a first optimization of the potentials needed to effectively move the charge towards the sense node, the overall yield of the electron counting capability of the Skipper-CCDs was quantified (71\%). Figure~\ref{fig:epeaks} (left) shows the yield per amplifier and buried-channel implantation dose. Sensors with $1\times10^{12}$~ions/cm$^{2}$ show the best performance in all amplifiers. 

\begin{figure}[h!]
    \centering
    \begin{subfigure}{}
    \includegraphics[height=3.5cm]{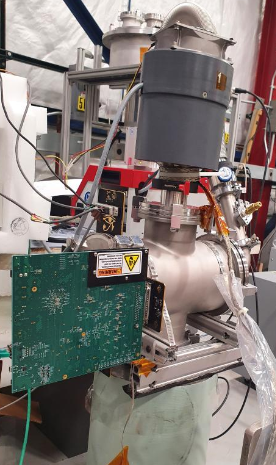}
    \end{subfigure}
    \begin{subfigure}{}
    \includegraphics[height=3.5cm]{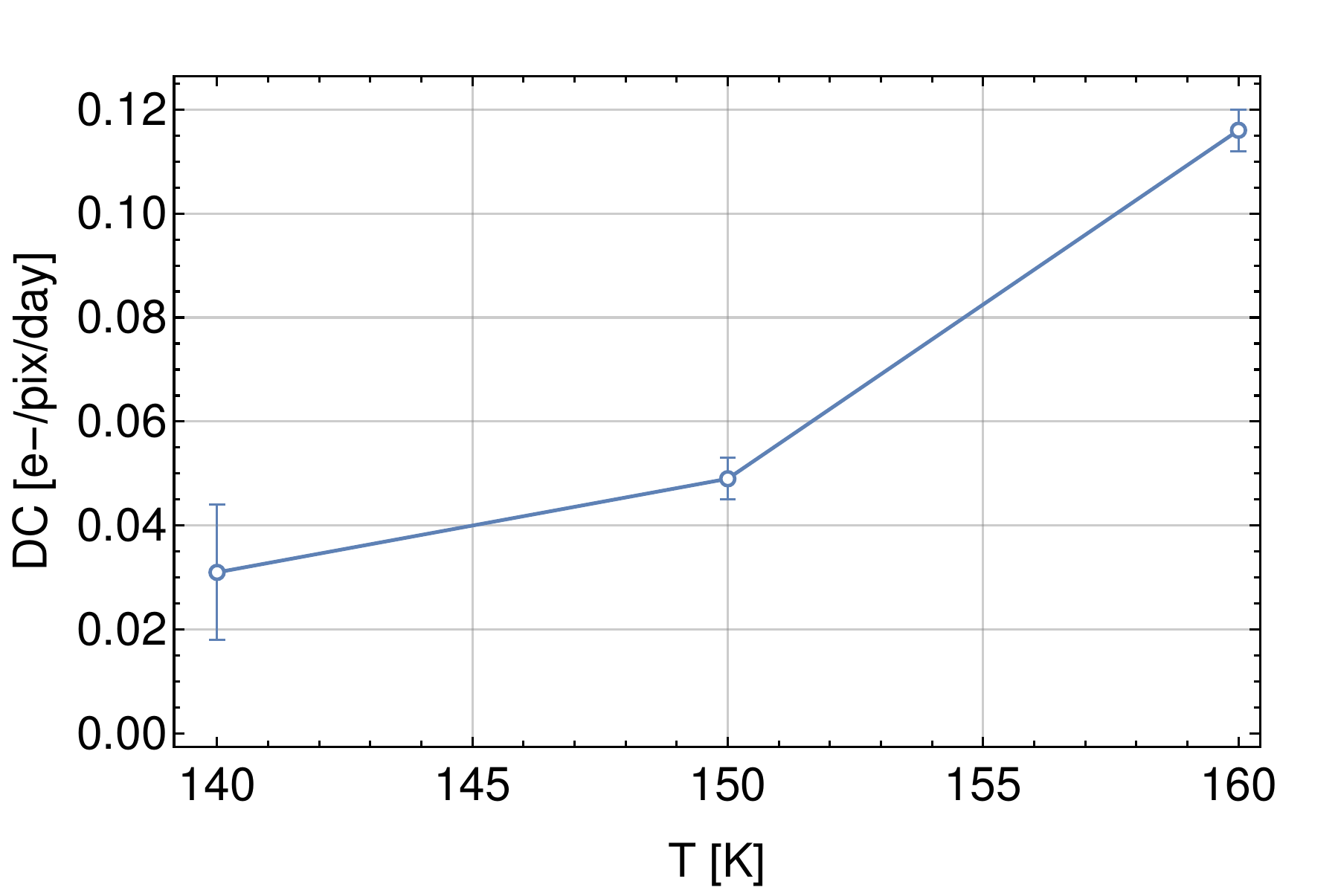}
    \end{subfigure}
    \caption{Left: Setup at FNAL with Pb shield. Right: DC vs T obtained from measurements using a Microchip Skipper-CCD installed in this setup.}
    \label{fig:DC}
\end{figure}

First measurements of the dark current (DC) of the Microchip Skipper-CCDs were done at different temperatures (Fig.~\ref{fig:DC}). Below 150~K this rate starts getting flat. This is contrary to what is theoretically expected if the main contribution to this rate was dark current, but similar to what has been seen in previous measurements with detectors operating above ground~\cite{Fernandez-Moroni2022}. The measured DC at 140~K is $(0.031\pm 0.013)$~e$^-$/pix/day. There are ongoing efforts to measure DC underground to determine if these sensors meet the Oscura requirements.

\section{Conclusions}
CCD technology has led to competitive results in astronomy and particle physics, e.g. DM~\cite{DAMIC2020,DAMICwithelectrons} and low-energy neutrinos~\cite{CONNIE2021,CONNIEmediators}. Skipper-CCDs allow the development of electron counting experiments with plenty of potential applications~\cite{Tiffenberg2017, SnowmassAstro2022, Fernandez-Moroni2020}. A direct DM search program is ongoing, with the 10~kg Oscura experiment being the ultimate goal. From the first fabrication run at Microchip, 71\% of the Oscura sensors reached sub-electron noise. We measured a DC at surface of $\sim0.031$~e$^-$/pix/day. These results demonstrate the success of the new fabrication process in 200~mm diameter wafers.

\bibliography{mybibfile}

\end{document}